# FEEDBACK BASED REPUTATION ON TOP OF THE BITCOIN BLOCKCHAIN

Davide Carboni

Information Society Department, CRS4, Loc. Piscina Manna, Pula, Italy
dcarboni@crs4.it

## ABSTRACT

*The ability to assess the reputation of a member in a web community is a need addressed in many different ways according to the many different stages in which the nature of communities has evolved over time. In the case of reputation of goods/services suppliers, the solutions available to prevent the feedback abuse are generally reliable but centralized under the control of few big Internet companies. In this paper we show how a decentralized and distributed feedback management system can be built on top of the Bitcoin blockchain.*

## 1. INTRODUCTION

The ability to assess the reputation of a member in a web community is an essential need that arose since the web was born. This concept took several different forms according to the different stages in which the nature of web communities and participation in digital life have evolved over time. Notable examples are: the rank of a forum member assigned according to the volume of messages and replies provided in systems like bulletin boards; the reputation gained by sellers and buyers in e-commerce communities like eBay which is in turn based on the feedback they provide each other after the conclusion of a commercial transaction. The last reification of the concept of "web community reputation" is related to social networks. This incarnation is now more pervasive than ever because the adoption of social networks is now mainstream among people of any kind, of any age, and of any cultural background. The number of like(s) in a Facebook page, or the followers of a Twitter profile are commonly accepted as success indicators of a person, a group, a project, or an organization. This success indicator is often considered an asset or even a credential to better position the popularity even in traditional media because TV programs and newspapers often relaunch news and gossip directly from social media monitoring the users that they consider trend makers. Even politicians have discovered the positive network effect of gaining visibility and thus reputation in social media up to the point that a market for fake-like or fake followers is now very profitable [1].

In the case of reputation of goods/services suppliers, another important factor comes into play and this factor is money.  So a web community like eBay must handle with great care the reputation score of sellers and find countermeasures against the false transactions and false identities created with the malicious purpose to falsely enhance the reputation of an actor by means of non-genuine feedback. The solutions implemented on major buyer/seller platforms to prevent this kind of abuse on the reputation mechanism can be assumed as reliable "enough". The meaning of "enough" comes to the fact that brokerage platforms have a big interest in increasing and maintaining their own reputation. A site like eBay has strong interest in implementing robust and fair reputation and trust mechanisms therefore tends to prohibit or limit any abuse. They have in principle no interest in falsifying the values of reputation of the participants and if they promote one of them despite its reputation they make it through some advertisement placement.

One practical and informal proof of the robustness of such feedback systems is in the fact they are still there and used by millions of users. It would therefore seem unnecessary to study new forms of feedback management beyond those implemented and deployed by big Internet companies, but we claim that is indeed necessary. The recent NSA scandal has shown how there has been excessive interference by national authorities in the operations of such big companies. It is likely that national interests of a dominant country could influence and control operations behind the scenes and may distort the reputations of participants of major trade/consumer platforms, for instance helping companies operating in one country at the expense of foreigner companies depending on the actual international politic conflicts. This is not a new concern, in the past it was discussed how a pervasive search engine like Google could be neutral and fair according to different geographical, political, racial interests [2][3]

In this paper we are going to illustrate that the enablers for decentralized feedback management already exist and are at once global, decentralized and secure. However, the main claim of this paper is that such enabling technologies not only exist, but they are already deployed and running, even if are actually not aimed at the specific objective of global reputation management.

In more analytic terms, we here assume the reputation as a numerical variable associated to an actor and, as compulsory requirement, its value is agreed by everyone and cannot be counterfeited by anyone. We can realize that these are, among the others, the same requirements of electronic money. Thus, given such equivalence, in a bijective proof, once we have a money management that is in turn global, decentralized, and secure we will also have a reputation management system with the same properties. Nowadays, we already have a system that meets the above requirements: the crypto-currency concept and Bitcoin is the most notable implementation. This digital currency is based on a chain of transactions replicated in thousands of nodes and unalterable thanks to state of the art cryptographic techniques.

Adopting the Bitcoin block chain for purposes different than money transfer is not a novel concept. For instance Namecoin[4] is a Bitcoin fork able to work both as money and as decentralized name service. In[5] is described a Bitcoin-based approach to evaluate trusted third parties among certification authorities built-in in user client software. In[6] is described a method to manage anonymous credentials based on the concept of "general distributed ledger" introduced by Bitcoin removing the need for a trusted credential issuer.

## 2. REPUTATION MODEL

In the field of incentive based feedback has been showed that incentive based mechanisms can have their equilibrium in honest reporting having truth-telling as a Nash equilibrium [7]. Unfortunately incentive based feedback can have also not truth-telling equilibrium, in [8] is reported a collusion-resistant system for incentive based reputation system. The model presented in our paper is not formally proved to be collusion resistant, but it is reasonably robust to be compared to feedback systems running on existing platform like eBay.

In the following sections, we illustrate how this model can be implemented on top of the Bitcoin protocol. As first step we define some concepts:

Service S: a service produced and consumed online. For instance a Web service API to transcode a video.

SLA Service Level Agreement, a digital document that describe what a service does and with which contractual obligations (e.g. get a response within a maximum time limit)

Consumer: the consumer of service S.

Producer: the producer of service S

Payment P: a money transaction, an amount is transferred from a participant to another. Here we assume a payment P is securely linked to a service S.

Voucher V: a voucher is transaction linked to a past payment P. A voucher must contain an amount of money called the vote fee equal to a percentage (e.g. 3%) of the payment P and optionally can contain an additional amount called the incentive.

The important concept behind the Voucher is that both parties must digitally sign it.

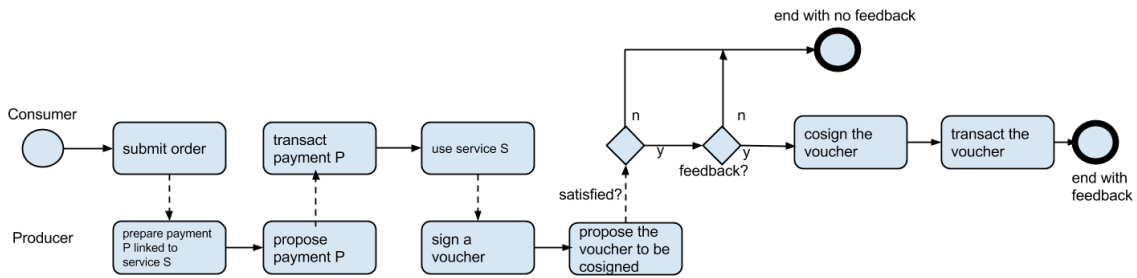

Fig.1 - workflow diagram. The consumer orders a service, performs the payment and if satisfied then decides whether or not to accept the incentive in the voucher leaving a positive feedback.

Having a voucher transaction completed is equivalent to logically increment the reputation for that service. The actual reputation for a service is equal to the sum of voting fees for that service.

## 3. SCENARIO

The BPM diagram for this scenario is visible in Fig. 1. Consumer pays 0.1 units to Producer for accessing 1000 non-concurrent invocations of a video transcoding API and each execution must complete within 5 minutes. The Producer builds also a Voucher transaction with a voting fee of 0.003 units and an incentive of 0.01 units and transmits the Voucher to Consumer to be co-signed.

If Producer complies with the contract then Consumer has two choices:

Consumer signs together with Producer the Voucher, this means validating the execution of the contract.

Consumer is not interested in the incentive. He does not co-sign the voucher.

In the case (1) the reputation of service S, and therefore the overall reputation of the Producer, increments by 0.003 (the voting fee). Otherwise (2) nothing happens to the actual reputation..

## 4. DISCUSSION ABOUT THE REPUTATION MODEL

The model proposed is quite simple, but in our opinion comparable with feedback systems actually implemented in main buyers/sellers community platforms sites. Here in this paper we are not proving the collusion-robustness of this model for incentive-based reputation management, but that such model even with its limitation, can be decentralized on existing Bitcoin infrastructure and running protocols without modification and without the need to fork the Bitcoin source tree to generate an alternative coin. This claim will be better explained in the rest of the paper. Here follows some qualitative properties and some issues related to the proposed model.

### 4.1. Sybil Attack

A producer may try to pump up his reputation using fake identities and assigning to them vouchers to vote. Given a voting fee for the transaction exists and is greater than zero, this attempt would result in expenditure for the producer more or less equivalent to advertise products. So he can climb the reputation slope not faster the amount of money he can afford. Thus the fraud is mitigated but this is still an issue, in fact reputation and advertisement are two different concepts that should not be mixed. The reputation should be a process incentivised but based on the sincere opinion of a consuming party about a product. A solution to further mitigate this attack is to calculate the reputation with a weighted sum, in other words gaining more reputation from peers that in turn have a good reputation and less from peers with low reputation.

### 4.2. Free Services

An opposite issue affects free services. According to the model presented in the previous section, if a service is free so happens that the voting fee would be zero and hence the reputation would always be zero. This is an undesirable side effect because even a free service should be able to have a reputation. To overcome this problem we may assume that a minimum fee could always be included also for free services, but here stands the dilemma of a service provider to afford such fees to gain reputation on a free service.

### 4.3. Mirror for the lark

Given that each voucher spent increases the service 's reputation by a fraction of the cost, a producer may gain high reputation with a few expensive products, and have an advantage with respect to competitors in selling low cost services. This really could happen only at the beginning of the life of one of the products, and then since it is possible to recalculate the reputation of every single service, then after a short amount of time it will be clear which products of a well-reputed producer are products with low reputation.

### 4.4. Unhappy but vote

A consumer may decide to take the incentive even if it is not really satisfied with the service S. He may eventually bring to arbitrage the case and try to get back the money of payment P. But as consequence of having signed and received the incentive he has de-facto validated the execution of the contract, thus his claim should not be taken into further consideration. The action of spending a voucher is a check-proof with timestamp and success acknowledgement. This consequence would discourage consumers from participating with the only objective to monetize the incentive.

## 5. ENABLING TECHNOLOGIES

Bitcoin protocol addresses the byzantine generals' problem [9]. In this problem the fact that every actor can write on a shared memory my produce an entangled set of conflicting assertions. The original problem has been formally proved to have no practical solution but in the subsequent works was conceived the concept of Byzantine fault tolerance and practical Byzantine fault tolerance [10]. Bitcoin is one of the modern implementation of the concept. It assumes that the cost of "writing" is not zero and the write access right is acquired after a proof-of-work. This proof-of-work is derived from the hashcash algorithm [11] and as final result, logically allows only one actor at time to "lock" the ledger and write. Details are available here [12]. The ledger in the Bitcoin terminology is a replicated log of transactions called the block chain. Transactions are collected in blocks and blocks are appended according to the proof-of-work algorithm mentioned above. In fig. 2 is showed a transaction.

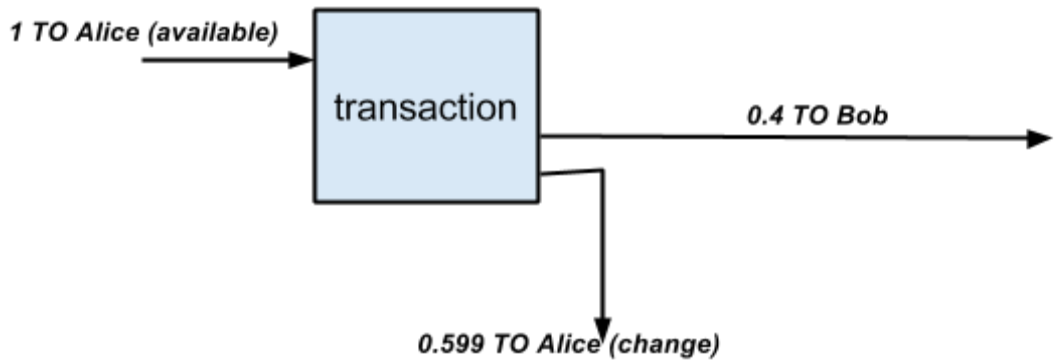

Fig. 2  The simplest Bitcoin transaction. In the example, an amount of 0.4 is transferred from an address "Alice" to an address "Bob". The input comes from a previous transaction to Alice. Notice that an amount of 0.599 is sent back to Alice as change. The 0.001 unassigned is an amount that the node solving the proof-of-work can assign to itself as transaction fee.

Each transaction is signed with the private key of the owner of a coin; in general a transaction may have many inputs, many outputs, and be co-signed by many private keys. In the simplest use case of one-to-one payments, a transaction has a single input and in general two output: the first output is to the Bitcoin address of the recipient (a public key), the second is the change and directed to the Bitcoin address of the sender, a small amount should be left unassigned and this will become the transaction fee.

In the next paragraph and Fig. 4 we show how it is possible to implement the use case described in the previous sections without adding any new feature to the existing Bitcoin protocol.

## 6. BITCOIN-BASED FEEDBACK SYSTEM

Lets now assume the service S is identified by its own URL, eg. S := http://foo.bar. Lets also redefine all the concepts already defined in previous sections but now enriched with implementation details directly related to the Bitcoin protocol. We define Payment P a money transaction in BTC where an amount is transferred from a Bitcoin address to another. Double spending is prevented by the nature of Bitcoin block chain. Here we need a way to cryptographically link the payment P to a service S. This link is realized generating a correct Bitcoin address starting from the hash of the URL of S and sending 0.0 BTC to such address. With this trick, the zero-valued monetary transaction is also bound to the service URL. A tool to check if a generated address is a valid Bitcoin address is available online at [13]. A similar approach is used in [14] which provides an online notary service to prove the existence of a document at a given time. To generate such syntax checked but fake Bitcoin address, we compute the ripemd160 [15] hash, a stronger version of RIPEMD, of S. The steps to generate a valid Bitcoin addres tied to a free form text are visible in Fig. 3. Lets call this address S*. Given that Bitcoin addresses contain a built-in check code; it's generally not possible to send Bitcoins to a mistyped or malformed address [16].

Lets now define the Voucher V as a transaction linked to a past payment P. A voucher must contain an amount of money, called the vote fee, equal to a percentage (e.g. 3%) of the payment P and optionally can contain an additional amount called the incentive. The important concept behind the Voucher is that both parties (the buyer and the seller) must sign it digitally. Only when this condition is met, the Voucher will become a valid transaction eligible to be inserted in the block chain. The vote fee is merely the amount of the voucher not directly connected to a recipient, as in Bitcoin, if a coin has a part unspent this part become "transaction fee" and will be taken by the fastest node solving the current proof-of-work. The probability that the producer will be also the node getting the voting fee is assumed negligible in a populated network. As side effect of having a "voting fee", quite palatable if compared with normal Bitcoin transaction fees, is that nodes involved in proof-of-work are also incentivised to seek such reputation transactions as a valuable source of income. Having a voucher transaction completed is equivalent to logically increment the reputation for that service. Traversing the block chain is possible to reconstruct the actual reputation for a service, which is equal to the sum of voting fees for those services.

Summarizing (see Fig. 4), the consumer builds and signs a transaction P with 3 outputs:

- Out1 is the cost of the service (0.1BTC) and directed to the producer
- Out2 is 0.0 and directed to S*
- Out3 is the change and redirected to the consumer

```
3 - RIPEMD-160 Hash of 2
925838B105839043883035691F3540A6829C9B68          [Send]

4 - Adding network bytes to 3
00925838B105839043883035691F3540A6829C9B68        [Send]

5 - SHA-256 hash of 4
CACAD5C48F7BCE15E80E9FB28255A56164DB720D3EB3057868177A8DC9DB81FB

6 - SHA-256 hash of 5
35764DBD8652FF31A08B8717086657E77944BFD1F6D1D90780BD6CF04413AF56

7 - First four bytes of 6
35764DBD

8 - Adding 7 at the end of 4
00925838B105839043883035691F3540A6829C9B6835764DBD    [Send]

9 - Base58 encoding of 8
1ELoLjz9smCcf8vyL9T2dPnp1iY7RmF61S                [Send]
```

Fig. 3 In this figure, the (9) is a valid Bitcoin address generated starting from a RIPEMD160 hash (3) of the URL http://foo.bar

Notice that the Out2 is useless as money transfer but it is a trick that allows to cryptographically establish a link between the transaction P and the service S. Moreover, if the sum directed to Out2 was greater than zero, this coin would be permanently destroyed from the p2p network because the recipient address is a formally valid address but no one has the private key to spend that coin. As illustrated in Fig. 4, the payment transaction cryptographically linked to the hash of serviceID, together with the voucher transaction signed by both parties[1], is the combination of Bitcoin transaction that marks the increase of actual reputation score for the producer.

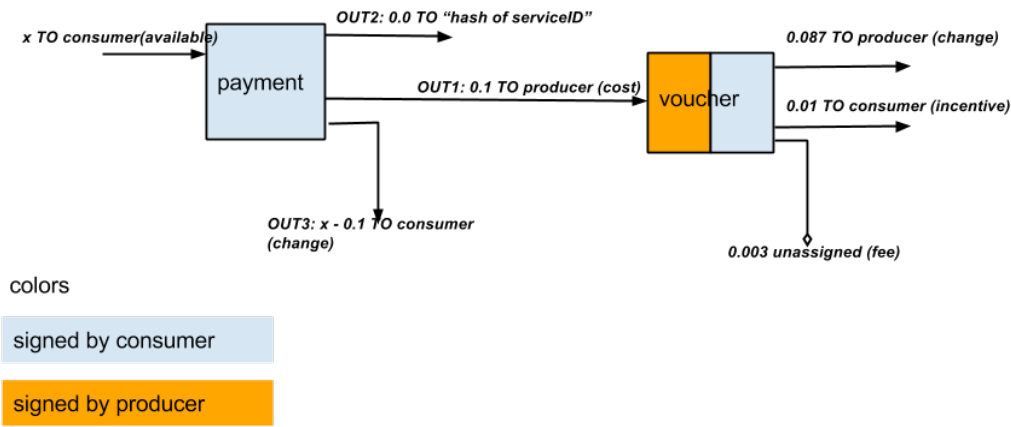

Fig. 4 – On the left side the payment transaction cryptographically linked to the hash of serviceID, on the right the voucher transaction signed by both parties and used to increase the actual reputation value for the producer.

Notice that the computation of the actual reputation is not part of the block chain protocol; it can be performed by third party applications that traverse the block chain and may provide different formulas to compute the reputation

## 7. REQUIREMENT FOR A DECENTRALIZED REPUTATION MANAGEMENT SYSTEM

In [17] are enumerated a list of requirements for a distributed and decentralized reputation management system. We decided to adopt these requirements as a benchmark of our reputation model. Here follows a discussion about each of the mentioned requirements.

> "The system should be self-policing. Policies are defined and enforced by the peers themselves and not by some central authority."

Having the system based on a crypto currency, it is not vulnerable to single point of failure. This assertion is as true as the proof-of-work capacity is distributed throughout the network. Having a single powerful actor able to close an important fraction of the total transaction blocks may cause the network to be attacked forging for instance a secondary head in the block chain.

---

[1] Technically speaking, a multisigned transaction is a transaction from a multisigned address. In this particular case would be a 2 out of 2 multisigned address. Details are not explained here for sake of clarity

Nevertheless, the Bitcoin protocol is a money transaction protocol and in such cases the real damage would not be in the falsified reputation of parties, but in the destruction of the monetary value of the coins because some actors will become able to perform double spending, and finally in the collapse of all the Bitcoin network. The informal proof is that the block chain still stands after four year of operations.

> "The system should maintain anonymity"

Anonymity is not a must-have requirement for the Bitcoin protocol. In fact, a Bitcoin transaction may be considered as anonymous as exchanging physical notes in a crowded square. The block chain is a "public place", inside every transaction it is possible to locate the Bitcoin address of whom receives a coin. The address is usually in a form like:

15VjRaDX3zpbA8PVnbrCAFzrVxN7ixHQ2C

Thus in principle it is like an unknown face in a public square with the main difference that everyone can take this address and can try to harvest some information using web search engines. For instance, if Alice uses this address to receive donations in a web forum then it would be easy for an attacker to associate a real identity to a Bitcoin address. Thus, the anonymity is guaranteed as long as users does not leak relevant information that can be used to associate their identity to a Bitcoin address.

> " The system should not assign any profit to newcomers. That is, reputation should be obtained by consistent good behavior through several transactions, and it should not be advantageous for malicious peers with poor reputations to continuously change their opaque identifiers to obtain newcomers status."

In the model presented in this paper, the reputation starts with a value equal to zero and must be positive. The reputation is managed like money with the difference it is not possible to transfer a reputation to another actor. In the scenario described in the previous sections, the reputation is somehow bound to a URL. In such cases, if the Internet domain is transferred from an actor to another, also his reputation is transferred.

> "The system should have minimal overhead in terms of computation, infrastructure, storage, and message complexity."

The proposed approach has a minimal overhead because it is based on a distributed, decentralized infrastructure already running for monetary transactions.

> " The system should be robust to malicious collectives of peers who know one another and attempt to collectively subvert the system"

This feature comes directly from the Bitcoin design. If we assume a small community as a minor part of the whole p2p network, they have some chance to operate a malicious behavior. In the paper [18] is described a possible attack, nevertheless this does not affect the validity of transaction but would affect the confidence of peers about the fairness of block solving race.

## 8. CONCLUSIONS AND OPEN ISSUES

In this paper we present how an incentive based feedback model can be implemented on top of Bitcoin protocols and on top of the existing Bitcoin network without modifications in its core operations. Throughout the paper, we introduce the incentive mechanism assumed in this work and we discuss pros and cons of the model. Some notions required to understand how the transactions are managed in the Bitcoin protocol are also introduced to finally expose the main contribution of the paper which is to describe how the incentive based feedback is directly implementable on top of the Bitcoin network.

The advantage of using a decentralized architecture without single point of control is to provide a feedback management system, which is in turn decentralized, secure and global. The proposed solution is analyzed against the requirements exposed in [17] chosen as theoretical benchmark for a decentralized reputation management system.

As stated in the paper, the proposed approach has some limitation that can be summarized as follows:

- it is possible for the malicious user to build fake identities and use them to pump his own reputation. This undesirable effect is strongly mitigated by the fact that it is expensive because real coins need to be spent but it is somehow still an issue.

- According to the model presented in the previous sections, if a service is free so happens that the voting fee would be zero and hence the reputation would always be zero. This is an undesirable side effect because even a free service should be able to have a reputation.

- The model presented is reasonably robust to be compared to existing online feedback systems but it is not formally proved to be collusion resistant.

As final note, the blockchain is a new way to enable applications on top of cryptographic techniques. Bitcoin based payments is just the first and most notable application, but a lot more are likely to come. Currently, the image of Bitcoin as a currency is suffering some troubles because some important frauds and illicit behaviours have taken places during its recent and short life. The most notable is the recent bankrupcy of the MtGox exchange [19] which was the biggest and most visited site to buy/sell Bitcoins against fiat currencies. It may seem a paradox to design a reputation system on top of a technology that is experiencing a general lack of trust, but we claim that the technological advancement of the Bitcoin protocol are real, and even if his history will be sporadically affected by negative facts, the Bitcoin cannot be de-invented and it will likely pave the way to a new class of application. In this respect Bitcoin has been defined "not the money of internet but the internet of money"[20].